\documentclass[reprint,onecolumn,showpacs,preprintnumbers,nofootinbib,amsmath,amssymb,aps,prd,floatfix,notitlepage]{revtex4-1}
\pdfoutput=1
\usepackage{graphicx}
\usepackage{bm}
\usepackage{subfigure}
\usepackage{url}
\usepackage[hyperindex]{hyperref}
\usepackage{color}
\usepackage[ddmmyy,24hr]{datetime}
\usepackage{bigdelim}
\usepackage{booktabs}
\usepackage{dcolumn}
\usepackage{multirow}
\usepackage{subfigure}
\usepackage{cancel}
\usepackage{stackrel}
\usepackage{paralist}
\usepackage{xspace}
\usepackage{slashed}
\newcommand{\nua}[1]{\ensuremath{\rlap{\kern-2.5pt\ensuremath{\overset{\scriptscriptstyle(-)}{\phantom{\nu}}}}{\ensuremath{{\nu}_{#1}}}}}

\usepackage{enumitem}

\makeatletter
\def\namedlabel#1#2{\begingroup
    #2%
    \def\@currentlabel{#2}%
    \phantomsection\label{#1}\endgroup
}
\makeatother

\begin{document}

\title{Short-baseline neutrino oscillations with 3+1 non-unitary mixing}

\author{C. Giunti}
\email{carlo.giunti@to.infn.it}
\affiliation{Istituto Nazionale di Fisica Nucleare (INFN), Sezione di Torino, Via P. Giuria 1, I--10125 Torino, Italy}

%\date{\dayofweekname{\day}{\month}{\year} \ddmmyydate\today, \currenttime}
\date{3 April 2019}

\begin{abstract}
We consider a scenario with unitary mixing of the three light standard neutrinos
and a non-unitary mixing contribution of a heavier massive neutrino
that can generate short-baseline neutrino oscillations.
We show that this scenario predicts constant flavor-changing probabilities
at short-baseline distances.
Therefore, it cannot explain the spectral distortions
observed in the
LSND and MiniBooNE appearance experiments.
On the other hand, the survival probabilities oscillate as functions of $L/E$
and could explain oscillations in short-baseline disappearance experiments.
We also derive the bounds on the mixing parameters from the
existing short-baseline neutrino oscillation data.
\end{abstract}

%\pacs{}

\maketitle

%\section{}
%\label{}

Neutrino oscillation experiments have established that
neutrinos are massive and mixed particles
(see Refs.~\cite{Giunti:2007ry,Bilenky:2018hbz}).
The results of solar, atmospheric and long-baseline
neutrino oscillation experiments are well fitted in the
minimal framework of three-neutrino mixing
\cite{deSalas:2017kay,Capozzi:2018ubv,Esteban:2018azc}.
However,
there are indications in favor of short-baseline oscillations that
require the addition of at least a fourth massive neutrino,
that is usually called sterile
(see the reviews in Refs.~\cite{Bilenky:1998dt,Maltoni:2004ei,GonzalezGarcia:2007ib,Conrad:2012qt,Gariazzo:2015rra,Giunti:2019aiy}).
This is the 3+1 neutrino mixing scenario,
with the flavor neutrino states given by the following
superpositions of four massive neutrino states:
\begin{equation}
| \nu_{\alpha} \rangle
=
\sum_{k=1}^{4}
U_{\alpha k}^{*}
\,
| \nu_{k} \rangle
,
\quad
\text{with}
\quad
\alpha=e,\mu,\tau,s
,
\label{mix1}
\end{equation}
where $U$ is a $4\times4$ mixing matrix.
For ultrarelativistic neutrinos with energy $E$,
the corresponding probability of $\nu_{\alpha}\to\nu_{\beta}$ oscillations
at a distance $L$ from the source is given by
\begin{equation}
P_{\alpha\beta}(L/E)
=
\left|
\sum_{k=1}^{4}
U_{\alpha k}^{*}
U_{\beta k}
\exp\!\left(
- i \dfrac{ \Delta{m}^2_{k1} L }{ 2 E }
\right)
\right|^2
,
\label{p1}
\end{equation}
where
$ \Delta{m}^2_{kj} \equiv m_{k}^2 - m_{j}^2 $.
In short-baseline neutrino oscillation experiments the
effects of the small solar and atmospheric squared-mass differences
$\Delta{m}^2_{21}$ and $\Delta{m}^2_{31}$
that generate oscillations at larger distances are negligible
and the effective oscillation probabilities
are given by
\begin{equation}
P_{\alpha\beta}^{\text{SBL}}(L/E)
=
\left|
\sum_{k=1}^{3}
U_{\alpha k}^{*}
U_{\beta k}
+
U_{\alpha 4}^{*}
U_{\beta 4}
\exp\!\left(
- i \dfrac{ \Delta{m}^2_{41} L }{ 2 E }
\right)
\right|^2
.
\label{p2}
\end{equation}
In the standard 3+1 mixing scheme the neutrino mixing is generated by
the unitary diagonalization of the neutrino mass matrix
(see Refs.~\cite{Giunti:2007ry,Bilenky:2018hbz}).
Therefore, the $4\times4$ mixing matrix $U$ is unitary,
with the unitarity relations
\begin{align}
\sum_{k=1}^{4}
U_{\alpha k}^{*}
U_{\beta k}
=
\delta_{\alpha\beta}
\null & \null
\quad
\text{for}
\quad
\alpha,\beta=e,\mu,\tau,s
,
\label{u41}
\\
\sum_{\alpha=e,\mu,\tau,s}
U_{\alpha k}^{*}
U_{\alpha j}
=
\delta_{kj}
\null & \null
\quad
\text{for}
\quad
k,j=1,2,3,4
.
\label{u42}
\end{align}
Taking into account the unitarity relation (\ref{u41}),
the short-baseline oscillation probabilities are given by
\begin{equation}
P_{\alpha\beta}^{\text{SBL}}(L/E)
=
\left|
\delta_{\alpha\beta}
-
U_{\alpha 4}^{*}
U_{\beta 4}
\left[
1
-
\exp\!\left(
- i \dfrac{ \Delta{m}^2_{41} L }{ 2 E }
\right)
\right]
\right|^2
,
\label{sp1}
\end{equation}
that can be written as
\cite{Bilenky:1996rw}
\begin{equation}
P_{\alpha\beta}^{\text{SBL}}(L/E)
=
\delta_{\alpha\beta}
-
4
|U_{\alpha 4}|^2
\left( \delta_{\alpha\beta} - |U_{\beta 4}|^2 \right)
\sin^2\!\left(
\dfrac{ \Delta{m}^2_{41} L }{ 4 E }
\right)
.
\label{sp2}
\end{equation}
This is the standard 3+1 formula that has been used in many phenomenological analyses of the data of
short-baseline neutrino oscillation experiments
(see the reviews in Refs.~\cite{Bilenky:1998dt,Maltoni:2004ei,GonzalezGarcia:2007ib,Conrad:2012qt,Gariazzo:2015rra,Giunti:2019aiy}).
The latest results on the values of the mixing parameters
$|U_{\alpha 4}|^2$ for $\alpha=e,\mu,\tau$
and
$\Delta{m}^2_{41}$
are given in Refs.~\cite{Gariazzo:2018mwd,Dentler:2018sju,Giunti:2019aiy}.

Recently, the authors of Ref.~\cite{Kim:2018uht}
considered the possibility that the $4\times4$ mixing matrix $U$ is not unitary,
keeping the unitarity of the $3\times3$ submatrix
of three-neutrino mixing that connects the three active flavor neutrinos
$\nu_{e}$,
$\nu_{\mu}$, and
$\nu_{\tau}$
to the three light massive neutrinos
$\nu_{1}$,
$\nu_{2}$, and
$\nu_{3}$,
i.e.
\begin{align}
\sum_{k=1}^{3}
U_{\alpha k}^{*}
U_{\beta k}
=
\delta_{\alpha\beta}
\null & \null
\quad
\text{for}
\quad
\alpha,\beta=e,\mu,\tau
,
\label{u31}
\\
\sum_{\alpha=e,\mu,\tau}
U_{\alpha k}^{*}
U_{\alpha j}
=
\delta_{kj}
\null & \null
\quad
\text{for}
\quad
k,j=1,2,3
.
\label{u32}
\end{align}
We refer to this scenario as 3+1 non-unitary mixing (NUM).
In this case the $3\times3$ mixing is due to the diagonalization of the
mass matrix of the three active flavor neutrinos
that is assumed to be generated by a mechanism that is different
from the mechanism that generates the mass of the heavier neutrino $\nu_{4}$.
The additional mixing of the flavor neutrinos with $\nu_{4}$ can be generated by non-standard interactions between $\nu_{4}$ and the charged leptons
$e$,
$\mu$, and
$\tau$
\cite{Cvetic:2017gkt}.
The elements $U_{\alpha 4}$ of the mixing matrix for $\alpha=e,\mu,\tau$
are the coefficients that quantify the non-standard interactions of $\nu_{4}$
with respect to the standard charged-current weak interaction.
Since in this scenario $\nu_{4}$ is interacting, we do not call it sterile.

In principle, the quantities $|U_{\alpha 4}|$
can have any value, smaller or bigger than one,
but the current bounds on non-standard effects
imply that they are very small,
as shown in Ref.~\cite{Kim:2018uht} from the analysis of non-oscillation data.
In the following we will derive their bounds from
short-baseline neutrino oscillation data.

Since the three light neutrinos take part in weak interactions
through the unitary $3\times3$ submatrix
and the new massive neutrino $\nu_{4}$ is observable
only through its interactions with the three charged leptons,
it is not useful to consider a $4\times4$ mixing matrix
(the fourth row is not well defined),
but only the $3\times4$ mixing matrix that connects the four massive neutrinos
to the three observable charged leptons:
\begin{equation}
U
=
\begin{pmatrix}
U_{e1} & U_{e2} & U_{e3} & U_{e4}
\\
U_{\mu1} & U_{\mu2} & U_{\mu3} & U_{\mu4}
\\
U_{\tau1} & U_{\tau2} & U_{\tau3} & U_{\tau4}
\end{pmatrix}
.
\label{mix3x4}
\end{equation}

The effective oscillation probabilities
in short-baseline experiments
in the 3+1 NUM scenario
can be obtained from Eq.~(\ref{p2}) with the constraint (\ref{u31}):
\begin{equation}
\widetilde{P}_{\alpha\beta}^{\text{SBL}}(L/E)
=
\delta_{\alpha\beta}
\left[
1
+
2
|U_{\alpha 4}|^2
\cos\!\left(
\dfrac{ \Delta{m}^2_{41} L }{ 2 E }
\right)
\right]
+
|U_{\alpha 4}|^2
|U_{\beta 4}|^2
.
\label{np2}
\end{equation}
Here and in the following we indicate with a tilde the probabilities
obtained in the 3+1 NUM scenario.
Note that the probabilities (\ref{np2})
depend only on the absolute values of the elements of the fourth column of the mixing matrix
and are the same for neutrinos and antineutrinos,
as in the standard unitary 3+1 scenario.
Apart from this similarity,
one can see immediately that the oscillation probabilities $\widetilde{P}_{\alpha\beta}^{\text{SBL}}(L/E)$
are different from the standard unitary 3+1 oscillation probabilities
$P_{\alpha\beta}^{\text{SBL}}(L/E)$
in Eq.~(\ref{sp2}).
Moreover,
the effective short-baseline probabilities $\widetilde{P}_{\alpha\beta}^{\text{SBL}}(L/E)$,
and in general all the probabilities
obtained from Eq.~(\ref{p1}) with the constraint (\ref{u31}),
are not true probabilities,
because some of them can be larger than one and
the sum of the probabilities over the flavor index do not add up to one.
Indeed,
from Eq.~(\ref{p1}) we have
\begin{equation}
\sum_{\beta=e,\mu,\tau,s}
P_{\alpha\beta}(L/E)
=
\sum_{k=1}^{4}
\sum_{j=1}^{4}
U_{\alpha k}^{*}
U_{\alpha j}
\exp\!\left(
- i \dfrac{ \Delta{m}^2_{kj} L }{ 2 E }
\right)
\sum_{\beta=e,\mu,\tau,s}
U_{\beta k}
U_{\beta j}^{*}
.
\label{unitsum}
\end{equation}
Using the $4\times4$ unitarity relations (\ref{u41}) and (\ref{u42}),
in the standard 3+1 mixing scheme
we obtain the normal probability sum rule
\begin{equation}
\sum_{\beta=e,\mu,\tau,s}
P_{\alpha\beta}(L/E)
=
1
.
\label{unitsumst}
\end{equation}
On the other hand,
in the 3+1 NUM scenario,
considering only the sum over the three active flavors
we obtain a quantity that depends on $L/E$ and has a maximum that is bigger than one:
\begin{equation}
\sum_{\beta=e,\mu,\tau}
\widetilde{P}_{\alpha\beta}(L/E)
=
1
+
|U_{\alpha 4}|^2
\sum_{\beta=e,\mu,\tau}
|U_{\beta 4}|^2
+
2 \operatorname{Re}\! \left[
\sum_{\beta=e,\mu,\tau}
U_{\alpha 4}^{*}
U_{\beta 4}
\sum_{j=1}^{3}
U_{\alpha j}
U_{\beta j}^{*}
\exp\!\left(
- i \dfrac{ \Delta{m}^2_{4j} L }{ 2 E }
\right)
\right]
,
\label{unitsumnonu}
\end{equation}
with
\begin{equation}
\operatorname{max}\!\left[
\sum_{\beta=e,\mu,\tau}
\widetilde{P}_{\alpha\beta}(L/E)
\right]
=
1 + |U_{\alpha 4}|^2
\left(
2 + \sum_{\beta=e,\mu,\tau} |U_{\beta 4}|^2
\right)
.
\label{unitsumnonumax}
\end{equation}

Other remarkable features of the 3+1 NUM scenario are that
the initial survival probabilities are larger than one
and
there are zero-distance flavor violations due to the non-orthogonality of the flavor neutrino states:
\begin{equation}
\widetilde{P}_{\alpha\beta}(0)
=
\delta_{\alpha\beta}
\left( 1 + |U_{\alpha 4}|^2 \right)^2
+
\left( 1 - \delta_{\alpha\beta} \right)
|U_{\alpha 4}|^2
|U_{\beta 4}|^2
\quad
\text{for}
\quad
\alpha,\beta=e,\mu,\tau
.
\label{p0}
\end{equation}
These features are due to the non-standard interactions that
generate the coupling of the flavor neutrinos with $\nu_{4}$,
and cause an increase by $1 + |U_{\alpha 4}|^2$
of the production and detection probabilities of $\nu_{\alpha}$
and
a probability $ |U_{\alpha 4}|^2 |U_{\beta 4}|^2 $
to detect a neutrino produced as $\nu_{\alpha}$ (i.e. with a charged lepton $\alpha$)
with a different flavor $\beta$
(i.e. observing a charged lepton $\beta$)
at zero distance.
These phenomena are analogous to those that have been studied
in the framework of non-standard interactions of the three light neutrinos
(see the review in Ref.~\cite{Ohlsson:2012kf}).

Turning back to the discussion of short-baseline oscillations,
for $\nua{\alpha}\to\nua{\beta}$ appearance experiments
with $\alpha\neq\beta$
we have
\begin{equation}
\widetilde{P}_{\alpha\neq\beta}^{\text{SBL}}
=
|U_{\alpha 4}|^2
|U_{\beta 4}|^2
,
\label{npapp}
\end{equation}
that is a constant not dependent on energy and distance,
and obviously coincides with the zero-distance flavor transition probability in Eq.~(\ref{p0}).
On the other hand,
one can see from Eq.~(\ref{np2}) that the survival probabilities
oscillate as functions of $L/E$.
This apparently strange behavior can be understood as follows.
Let us write the flavor state in Eq.~(\ref{mix1}) as
\begin{equation}
| \nu_{\alpha} \rangle
=
| \nu_{\alpha}^{\text{W}} \rangle
+
U_{\alpha 4}^{*}
\,
| \nu_{4} \rangle
,
\quad
\text{with}
\quad
| \nu_{\alpha}^{\text{W}} \rangle
=
\sum_{k=1}^{3}
U_{\alpha k}^{*}
\,
| \nu_{k} \rangle
,
\quad
\text{for}
\quad
\alpha=e,\mu,\tau
.
\label{mix2}
\end{equation}
The states
$| \nu_{\alpha}^{\text{W}} \rangle$
are the standard weak-interaction states,
that are orthonormal because of the assumed unitarity of the $3\times3$ submatrix
of three-neutrino mixing
($ \langle \nu_{\alpha}^{\text{W}} | \nu_{\beta}^{\text{W}} \rangle = \delta_{\alpha\beta} $).
We also have
$ \langle \nu_{\alpha}^{\text{W}} | \nu_{4} \rangle = 0 $
because of the orthonormality of the mass eigenstates.
At short-baseline distances an initial state
$| \nu_{\alpha} \rangle$
evolves to
\begin{equation}
| \nu_{\alpha}(t) \rangle_{\text{SBL}}
\simeq
e^{-i E t}
\,
| \nu_{\alpha}^{\text{W}} \rangle
+
e^{-i E_{4} t}
\,
U_{\alpha 4}^{*}
\,
| \nu_{4} \rangle
.
\label{evo}
\end{equation}
when we project this state on a different flavor state
$| \nu_{\beta} \rangle$ with $\beta\neq\alpha$
the standard weak-interaction states do not contribute because of their orthogonality
and we obtain a constant transition probability,
because there is no phase difference to generate interference.
On the other hand,
if we consider the survival probabilities,
when we project the state (\ref{evo}) on $| \nu_{\alpha} \rangle$
there are contributions both from the standard weak-interaction states
and $| \nu_{4} \rangle$,
with a phase difference that generates interference.

Since the short-baseline flavor transition probability (\ref{npapp})
does not depend on energy,
it cannot fit spectral distortions as those observed in the
LSND \cite{Aguilar:2001ty}
and
MiniBooNE \cite{Aguilar-Arevalo:2018gpe}
$\nua{\mu}\to\nua{e}$ appearance experiments.
Hence,
it cannot explain these anomalies.
This is, however, not a vital defect of the 3+1 NUM scenario with respect
to the standard unitary 3+1 mixing case,
because the standard unitary 3+1 interpretation of the current data suffers
of a strong tension between the appearance and disappearance data
mainly due to the incompatibility between the
MINOS\&MINOS+ bound on $\nua{\mu}$ disappearance \cite{Adamson:2017uda}
and the
LSND and MiniBooNE signals of $\nua{\mu}\to\nua{e}$ appearance
\cite{Gariazzo:2018mwd,Dentler:2018sju},
that disfavor the latter
(see the discussion in Ref.~\cite{Giunti:2019aiy}).

We can constrain the mixing matrix elements
in the 3+1 NUM scenario
by considering the exclusion curves
in the
$\sin^2\!2\vartheta_{\alpha\beta}$--$\Delta{m}^2_{41}$
plane
of short-baseline
$\nua{\alpha}\to\nua{\beta}$ appearance experiments.
At large values of $\Delta{m}^2_{41}$
these experiments are sensitive only to the constant averaged standard probability
\begin{equation}
\langle P_{\alpha\beta}^{\text{SBL}} \rangle
=
\frac{1}{2} \, \sin^2\!2\vartheta_{\alpha\beta}
.
\label{asp2}
\end{equation}
From Eqs.~(\ref{npapp}) and (\ref{asp2})
we obtain the limits
\begin{equation}
|U_{\alpha 4}|^2
|U_{\beta 4}|^2
<
\frac{1}{2}
\left.
\sin^2\!2\vartheta_{\alpha\beta}^{\text{max}}
\right|_{\text{large $\Delta{m}^2$}}
.
\label{lapp}
\end{equation}
The best limits are given by the results of the NOMAD experiment:
\begin{align}
|U_{\mu4}|^2
|U_{e4}|^2
<
7.0 \times 10^{-4} % 1.4 \times 10^{-3}
\null & \null
\quad
\text{(NOMAD \protect\cite{Astier:2003gs})}
,
\label{muel1}
\\
|U_{\mu4}|^2
|U_{\tau4}|^2
<
1.7 \times 10^{-4} % 3.3 \times 10^{-4}
\null & \null
\quad
\text{(NOMAD \protect\cite{Astier:2001yj})}
,
\label{muta1}
\\
|U_{e4}|^2
|U_{\tau4}|^2
<
7.5 \times 10^{-3} % 1.5 \times 10^{-2}
\null & \null
\quad
\text{(NOMAD \protect\cite{Astier:2001yj})}
,
\label{elta1}
\end{align}
at 90\% CL.
These limits are corroborated by the slightly less stringent
90\% CL limits
\begin{align}
|U_{\mu4}|^2
|U_{e4}|^2
<
8.5 \times 10^{-4} % 1.7 \times 10^{-3}
\null & \null
\quad
\text{(KARMEN \protect\cite{Armbruster:2002mp})}
,
\label{muel21}
\\
|U_{\mu4}|^2
|U_{e4}|^2
<
9.0 \times 10^{-4} % 1.8 \times 10^{-3}
\null & \null
\quad
\text{(CCFR \protect\cite{Romosan:1996nh})}
,
\label{muel22}
\\
|U_{\mu4}|^2
|U_{\tau4}|^2
<
2.2 \times 10^{-4} % 4.4 \times 10^{-4}
\null & \null
\quad
\text{(CHORUS \protect\cite{Eskut:2007rn})}
,
\label{muta2}
\\
|U_{e4}|^2
|U_{\tau4}|^2
<
2.2 \times 10^{-2} % 4.4 \times 10^{-2}
\null & \null
\quad
\text{(CHORUS \protect\cite{Eskut:2007rn})}
.
\label{elta2}
\end{align}
Therefore,
there are robust stringent limits on the products
$ |U_{\alpha 4}|^2 |U_{\beta 4}|^2 $
and the corresponding flavor appearance transitions
in the 3+1 NUM scenario.

Let us now consider short-baseline disappearance experiments.
From Eq.~(\ref{np2}),
the survival probabilities can be written as
\begin{equation}
\widetilde{P}_{\alpha\alpha}^{\text{SBL}}(L/E)
=
\left( 1 + |U_{\alpha 4}|^2 \right)^2
-
4
|U_{\alpha 4}|^2
\sin^2\!\left(
\dfrac{ \Delta{m}^2_{41} L }{ 4 E }
\right)
.
\label{npdis}
\end{equation}
These survival probabilities oscillate between the maximal value
$\left( 1 + |U_{\alpha 4}|^2 \right)^2$
that is larger than one
and the minimal value
$\left( 1 - |U_{\alpha 4}|^2 \right)^2$
that is smaller than one,
with the average
$1 + |U_{\alpha 4}|^4$
that is larger than one.
Since these survival probabilities have a form that is different from the standard ones
given by Eq.~(\ref{sp2}),
in general
it is not possible to constrain the quantities $|U_{\alpha 4}|^2$
using the results of the standard unitary 3+1 analyses of short-baseline disappearance experiments.
However,
there is an approximate correspondence for the analyses based on the
relative comparison of the survival probabilities at different distances
or different energies for small values of $|U_{\alpha 4}|^2$.
For example,
for the ratio of the probabilities measured with
a far detector at $L_{\text{F}}$
and
a near detector at $L_{\text{N}}$
we have
\begin{equation}
\dfrac{\widetilde{P}_{\alpha\alpha}^{\text{SBL}}(L_{\text{F}}/E)}{\widetilde{P}_{\alpha\alpha}^{\text{SBL}}(L_{\text{N}}/E)}
\simeq
\dfrac{P_{\alpha\alpha}^{\text{SBL}}(L_{\text{F}}/E)}{P_{\alpha\alpha}^{\text{SBL}}(L_{\text{N}}/E)}
=
1
-
4 |U_{\alpha 4}|^2
\left[
\sin^2\!\left(
\dfrac{ \Delta{m}^2_{41} L_{\text{F}} }{ 4 E }
\right)
-
\sin^2\!\left(
\dfrac{ \Delta{m}^2_{41} L_{\text{N}} }{ 4 E }
\right)
\right]
+
\text{O}(|U_{\alpha 4}|^4)
.
\label{rfn}
\end{equation}
For spectral distortion analyses, we can consider the ratio between the probability and its average:
\begin{equation}
\dfrac{\widetilde{P}_{\alpha\alpha}^{\text{SBL}}(L/E)}{\langle\widetilde{P}_{\alpha\alpha}^{\text{SBL}}\rangle}
\simeq
\dfrac{P_{\alpha\alpha}^{\text{SBL}}(L/E)}{\langle{P}_{\alpha\alpha}^{\text{SBL}}\rangle}
=
1 + 2 |U_{\alpha 4}|^2
-
4
|U_{\alpha 4}|^2
\sin^2\!\left(
\dfrac{ \Delta{m}^2_{41} L }{ 4 E }
\right)
+
\text{O}(|U_{\alpha 4}|^4)
.
\label{rsp}
\end{equation}
These approximate equalities allow us to constrain
small values of
$|U_{e4}|^2$ and $|U_{\mu4}|^2$
in the 3+1 NUM scenario using the results
of the standard unitary 3+1 analyses of several short-baseline disappearance experiments.

Let us consider first $\nu_{e}$ and $\bar\nu_{e}$ disappearance.
The results of the standard unitary 3+1 analyses
of the absolute rates of reactor neutrino experiments
that led to the reactor electron antineutrino anomaly
\cite{Mention:2011rk}
and
those of the Gallium radioactive source experiments
that led to the Gallium electron neutrino anomaly
\cite{Giunti:2010zu}
cannot be used in the 3+1 NUM scenario because of the different
normalization of the survival probability with respect to the standard one.
However,
the recent short-baseline reactor neutrino experiments obtained information on
$\bar\nu_{e}$ disappearance that is independent of the absolute neutrino flux
through the ratio of the spectra measured at different distances.
Therefore,
their results can be applied to the 3+1 NUM scenario through Eq.~(\ref{rfn})
for small values of $|U_{e4}|^2$.
In particular,
from the combined fit of the data of the
NEOS \cite{Ko:2016owz}
and
DANSS \cite{Alekseev:2018efk}
experiments
\begin{equation}
|U_{e4}|^2 = 0.012 \pm 0.003
\quad
\text{and}
\quad
\Delta{m}^2_{41} = 1.29 \pm 0.03 \, \text{eV}^2
\quad
\text{\protect\cite{Gariazzo:2018mwd}}
.
\label{NEOS+DANSS}
\end{equation}
Since the inclusion of other experimental data on short-baseline
$\nu_{e}$ and $\bar\nu_{e}$ disappearance
do not modify substantially the results in Eq.~(\ref{NEOS+DANSS})
in the standard scenario
\cite{Gariazzo:2018mwd,Dentler:2018sju},
we can consider the values in Eq.~(\ref{NEOS+DANSS})
as a good approximation of the global fit values
also in the 3+1 NUM scenario.

The value of $|U_{e4}|^2$ in Eq.~(\ref{NEOS+DANSS})
was considered also in Ref.~\cite{Kim:2018uht},
albeit without a correct derivation.
Using a correct derivation, we confirm the statement in Ref.~\cite{Kim:2018uht}
that the results of the fit of short-baseline neutrino oscillation data
is compatible with the value of $|U_{e4}|^2$ obtained in Ref.~\cite{Kim:2018uht}
from non-oscillation data
($|U_{e4}|^2 = 0.008 \pm 0.005$).

The positive indication (\ref{NEOS+DANSS}) in favor of
the existence of new physics given by
the combined analysis of the
NEOS
and
DANSS
data is under investigation in the ongoing
STEREO \cite{Almazan:2018wln},
PROSPECT \cite{Ashenfelter:2018iov},
SoLid \cite{Abreu:2018ajc}, and
Neutrino-4 \cite{Serebrov:2018vdw}
reactor experiments.
From a model-building point of view
the value of $|U_{e4}|^2$ in Eq.~(\ref{NEOS+DANSS})
is rather large and its experimental confirmation will require
dedicated theoretical investigations.
For example,
in the left-right symmetric models considered in Ref.~\cite{Cvetic:2017gkt}
$|U_{e4}|^2 \approx 0.012$ implies a value
$M_{R} \approx 250 \, \text{GeV}$
of the mass of the vector boson $W_{R}$ that mediates right-handed charged-current weak interactions.
This value of $M_{R}$ is smaller than the bound $M_{R} > 300 \, \text{GeV}$
found in a general analysis of left-right symmetric models \cite{Langacker:1989xa}
(as explained in Ref.~\cite{Cvetic:2017gkt},
the larger lower limits on $M_{R}$ obtained in analyses of LHC data
\cite{Maiezza:2010ic,CMS:2017ilm}
have been obtained in specific left-right models).

\begin{figure*}[!t]
\centering
\setlength{\tabcolsep}{0pt}
\begin{tabular}{cc}
\subfigure[]{\label{fig:um4-minos}
\includegraphics*[width=0.49\linewidth]{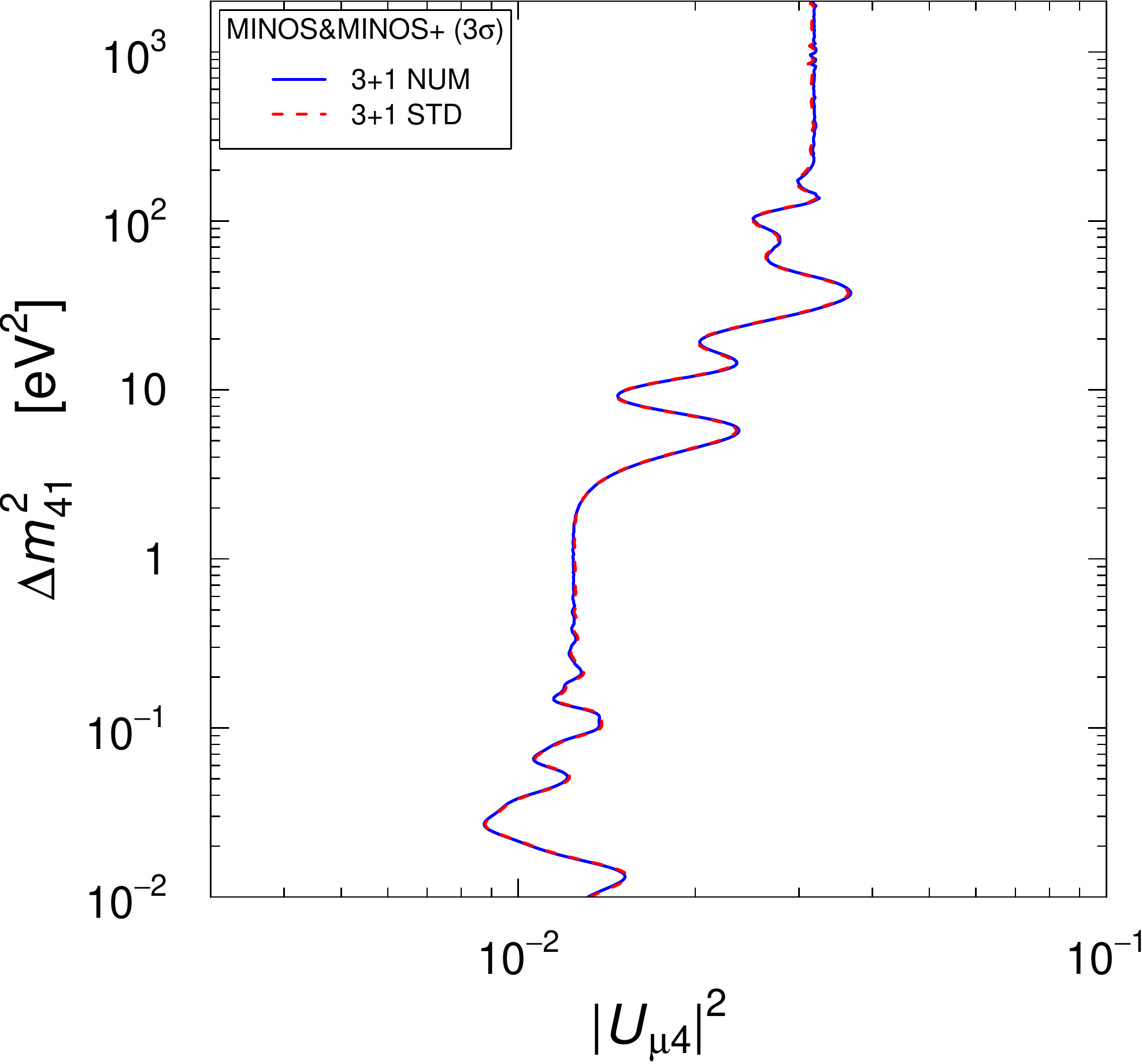}
}
&
\subfigure[]{\label{fig:um4-all}
\includegraphics*[width=0.49\linewidth]{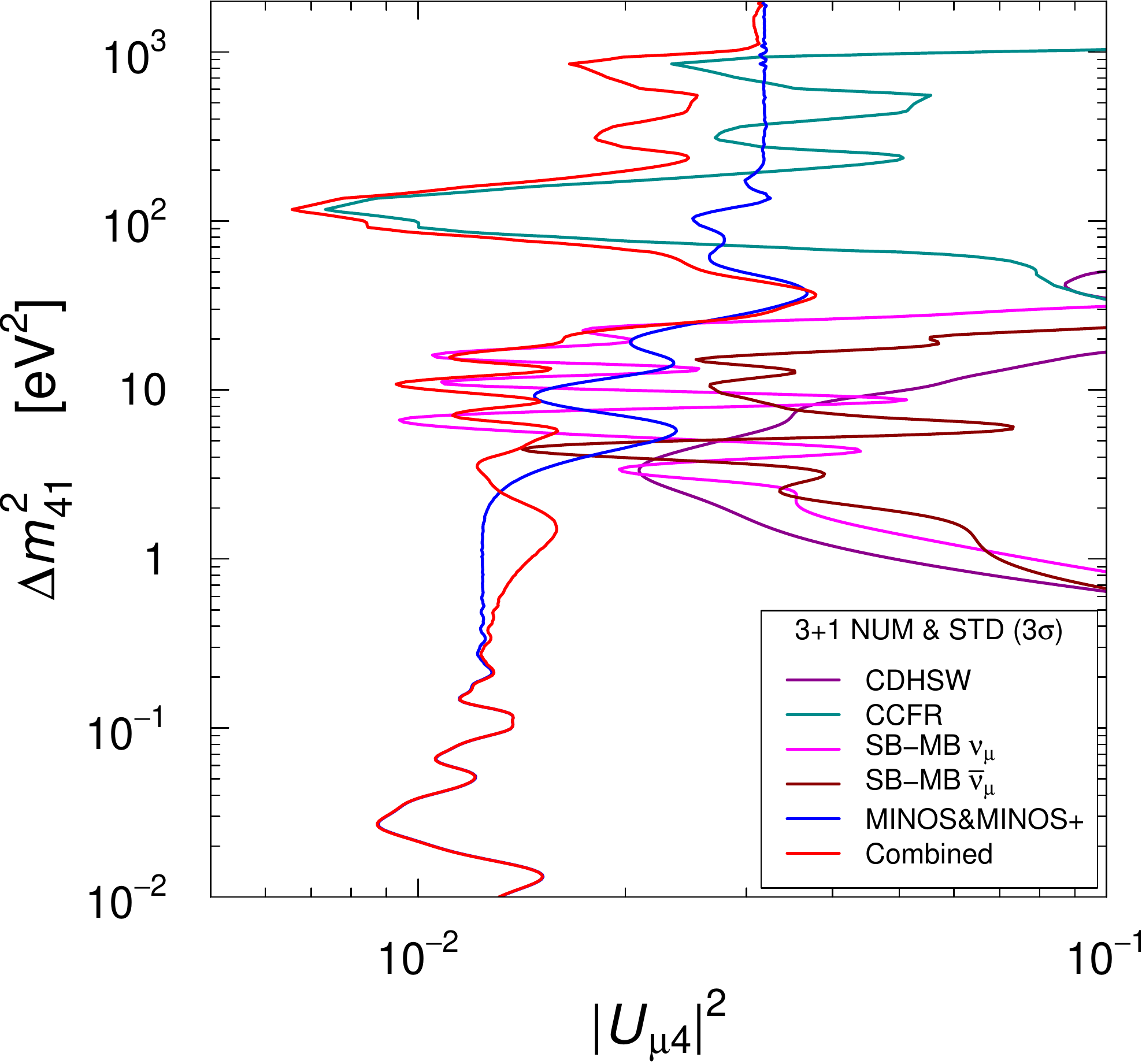}
}
\end{tabular}
\caption{ \label{fig:um4}
Exclusion curves in the
$|U_{\mu4}|^2$--$\Delta{m}^{2}_{41}$ plane.
\subref{fig:um4-minos}
Comparison of the $3\sigma$ exclusion curves
obtained from the MINOS\&MINOS+ data
\protect\cite{Adamson:2017uda}
in the 3+1 non-unitary mixing (NUM)
and 3+1 standard unitary mixing (STD) scenarios.
\subref{fig:um4-all}
Exclusion curves of the
CDHSW \protect\cite{Dydak:1983zq},
CCFR \protect\cite{Stockdale:1984cg},
SciBooNE-MiniBooNE \protect\cite{Mahn:2011ea,Cheng:2012yy}, and
MINOS\&MINOS+
in both the 3+1 non-unitary and standard unitary mixing scenarios
and the total combined exclusion curve.
}
\end{figure*}

Let us now consider short-baseline $\nu_{\mu}$ and $\bar\nu_{\mu}$ disappearance,
for which there is currently no experimental indication.
The limits on $|U_{\mu4}|^2$ in the standard unitary 3+1 scenario are dominated
by the MINOS\&MINOS+ bound \cite{Adamson:2017uda}
for $\Delta{m}^2_{41} \lesssim 3 \, \text{eV}^2$
and
$30 \lesssim \Delta{m}^2_{41} \lesssim 70 \, \text{eV}^2$
(see Ref.~\cite{Giunti:2019aiy}).
Since this bound was obtained taking into account the
oscillations in the near and far detectors
with a phenomenological neutrino flux constraint
and taking into account the oscillations
due to the atmospheric squared-mass difference in the far detector,
we need to reanalyze the MINOS\&MINOS+ data
in the 3+1 NUM scenario in order to check if there is a change of the limits on $|U_{\mu4}|^2$
with respect to the standard unitary 3+1 scenario.
We performed this analysis adapting the MINOS\&MINOS+
code that is available in the data release of Ref.~\cite{Adamson:2017uda}.
The result is depicted in Fig.~\ref{fig:um4-minos},
that shows that the limits on $|U_{\mu4}|^2$
in the NUM and standard unitary 3+1 scenarios are practically the same.
This is due to the smallness of $|U_{\mu4}|^2$
(we checked that the $\chi^2$ is different for large values of $|U_{\mu4}|^2$,
that are excluded at very large confidence levels).

Other stringent limits on $\nu_{\mu}$ and $\bar\nu_{\mu}$ disappearance
that are relevant in different ranges of
$\Delta{m}^2_{41}$
have been obtained in the
CDHSW \cite{Dydak:1983zq},
CCFR \cite{Stockdale:1984cg}, and
SciBooNE-MiniBooNE \cite{Mahn:2011ea,Cheng:2012yy}
experiments
through spectral ratio measurements at different distances.
Considering these limits for small values of $|U_{\mu4}|^2$,
we can apply the approximate equality (\ref{rfn})
and obtain the exclusion curves in Fig.~\ref{fig:um4-all}.

The total combined exclusion curve in Fig.~\ref{fig:um4-all}
shows that $|U_{\mu4}|^2$
is severely restricted for all values of $\Delta{m}^2_{41}$
larger than $10^{-2} \, \text{eV}^2$.
The small discrepancy between the combined limit and the dominating MINOS\&MINOS+ limit
for $\Delta{m}^2_{41} \sim 1 \, \text{eV}^2$
is due to a slight preference of the SciBooNE-MiniBooNE data for values of $|U_{\mu4}|^2$
that are larger than the MINOS\&MINOS+ limit.
This issue will be discussed in detail in Ref.~\cite{Gariazzo-Giunti-Ternes-IP-19}.

In conclusion, we have shown that the
expressions for the oscillation probabilities
in the 3+1 non-unitary mixing scenario are different from
those in the standard 3+1 unitary neutrino mixing case.
The phenomenology of short-baseline neutrino oscillations
have some differences in the two cases.
In particular,
in the 3+1 NUM scenario the flavor-changing probabilities are constant
at short-baseline distances and equal to their zero-distance value.
Therefore, 3+1 NUM cannot explain the spectral distortions
observed in the
LSND and MiniBooNE appearance experiments.
We presented in Eqs.~(\ref{muel1})--(\ref{elta2})
the current bounds on the non-unitary mixing from short-baseline appearance experiments.
We have also shown that the
short-baseline flavor survival probabilities
oscillate as functions of $L/E$ and the phenomenology of short-baseline
disappearance experiment based on the measurement of spectral distortions
or ratios of rates at different distances is approximately equal to that in
the standard 3+1 unitary mixing case for small values of the non-unitary mixing parameters.
This approximation implies the bounds on $|U_{e4}|^2$ and $|U_{\mu4}|^2$
presented, respectively, in Eq.~(\ref{NEOS+DANSS}) and Fig.~\ref{fig:um4-all}.

%\bibliographystyle{apsrev4-1}
%\bibliography{num}

\end{document}